\documentclass[12pt]{article}
\usepackage{graphicx}
\usepackage{epsfig}
\begin{document}
\begin{center}

{\bf PAIRING MATRIX ELEMENTS AND PAIRING GAPS WITH 
BARE, EFFECTIVE AND INDUCED INTERACTIONS}

\vspace{4ex}

F. Barranco$^1$, P.F. Bortignon$^{2,3}$,
R.A. Broglia$^{2,3,4}$, G. Col\`o$^{2,3}$, \\
P. Schuck$^5$, E. Vigezzi$^3$ and X. Vi\~{n}as$^6$ \\

\vspace{2ex}

$^1$ Departamento de Fisica Aplicada III, Escuela Superior de Ingenieros,\\
Universidad de Sevilla, Camino de los Descubrimientos s/n, 41092 Sevilla, Spain\\
$^2$ Dipartimento di Fisica, Universit\`a degli Studi, via Celoria 16,
20133 Milano, Italy\\
$^3$ INFN Sezione di Milano, via Celoria 16, 20133 Milano, Italy\\
$^4$ The Niels Bohr Institute, University of Copenhagen, Blegdamsvej 17,
20100 Copenhagen \O, Denmark\\
$^5$ Institut de Physique Nucl\'eaire, 15 rue Georges Cl\'emenceau,
91406 Orsay Cedex, France\\
$^6$ Departament d'Estructura i Constituents de la Mat\`eria,
Facultat de F\`{\i}sica,\\
Universitat de Barcelona, Diagonal 647, 08028 Barcelona, Spain

\vspace{1ex}

\begin{abstract}

The dependence on the single-particle states of the pairing matrix elements
of the Gogny force and of the bare low-momentum nucleon-nucleon potential
$v_{low-k}$  -
designed so as to reproduce the low-energy observables
avoiding the use of a repulsive core -  is studied in the semiclassical
approximation for the case of a typical finite, superfluid nucleus
($^{120}$Sn). It is found that the matrix elements of $v_{low-k}$ follow
closely those of $v_{Gogny}$ on a wide range of energy values 
around 
the Fermi energy $e_F$,
those associated with $v_{low-k}$ being less attractive. This
result explains the fact that around $e_F$ the pairing gap
$\Delta_{Gogny}$ associated with the Gogny interaction (and with a
density of single-particle levels corresponding to an effective 
$k$-mass $m_k
\approx 0.7 m$) is a factor of about 2 larger than $\Delta_{low-k}$,
being in agreement with $\Delta_{exp}$= 1.4 MeV. The exchange of low-lying 
collective surface vibrations among pairs of nucleons moving in time-reversal
states gives rise to an induced pairing interaction $v_{ind}$ peaked 
at $e_F$.
%and is different from zero
%in an energy range of 5 MeV around $e_F$. 
The interaction  $(v_{low-k}+ v_{ind})Z_{\omega}$ arising from 
the renormalization of the bare nucleon-nucleon potential and of the 
single-particle motion ($\omega-$mass and quasiparticle strength
$Z_{\omega}$) due to the particle-vibration coupling leads to a value
of the pairing gap at the Fermi energy $\Delta_{ren}$ which accounts for the
experimental value.

An important question which remains to be studied  quantitatively is 
to which extent $\Delta_{Gogny}$, which depends on average parameters,
and $\Delta_{ren}$, which explicitly depends on the parameters describing 
the (low-energy) nuclear structure, display or not a similar isotopic dependence,
and whether this dependence is born out by the data.

%While $\Delta_{Gogny}$ displays a smooth variation with mass number,
%$\Delta_{ren}$ contains all the elements to provide a detailed account of
%the isotopic dependence of the pairing gap, being sensitive to the detailed
%structure of each particular nucleus.

\end{abstract}

\end{center}

\vspace{1.5cm}

\section{Introduction}

An economic description of pairing correlations in finite nuclei is
provided by Hartree-Fock-Bogoliubov (HFB) theory~\cite{ripka,intro} 
making use
of phenomenological interactions like e.g. the finite range 
Gogny force~\cite{Decharge}
or density dependent zero-range forces combined with appropriate energy cut-offs
(cf., e.g.,~\cite{dd,dds}). Such a description leads to values of the pairing gap
which are in overall agreement with the experimental findings. Note
that in these calculations, the density of levels at the Fermi energy
$\rho(e_F)$ is
controlled by the so called $k-$mass (i.e. $\rho(e_F) \sim m_k$)
which, as a rule, is smaller than the bare
nucleon mass (e.g. $m_k \approx 0.7 m$ in the case of the Gogny force).

On the other hand, a number of studies have shown that the superfluid properties of nuclear
systems, ranging from infinite nuclear and neutron  matter to finite atomic
nuclei, are strongly influenced by polarization phenomena~\cite{effmass}.
In these calculations one starts from a bare nucleon-nucleon potential
(Argonne, Bonn, Paris, etc.) adding afterwards the renormalization
processes. Recently, the pairing gap, the quasiparticle spectrum
and the collective modes of $^{120}$Sn have been calculated solving the
Dyson-Gor'kov equation~\cite{epj}, in a single-particle
space characterized by $m_k = 0.7 m$, allowing the nucleons to interact
in the $^{1}S_0$ channel
through a $v_{14}$ Argonne N-N potential taking into account the variety
of renormalization processes (self-energy, fragmentation, induced interaction
and vertex corrections) arising from the coupling of the particles with
surface vibrations.  While the bare N-N interaction accounts for about half of the pairing gap,
overall agreement with the experimental findings is achieved by including
medium polarization effects.

In the present paper we want to shed light into the physics of these results,
by studying the magnitude of the different pairing ($J^\pi=0^+$) 
matrix elements
as well as their dependence on the energy of the single-particle states
lying in the vicinity of the Fermi energy in Section 2, and by discussing
the calculation of the pairing gaps associated with these matrix elements
in Section 3.

\section{Matrix elements}

The matrix element of the induced interaction can be written as
\cite{PRL}
\begin{equation}
 \langle \nu' \bar \nu' | v_{ind} | \nu,\bar \nu
  \rangle  =
   2 \sum_{LMn} \frac{<\nu|f_{Ln} Y_{LM}|\nu'><\bar\nu|f_{Ln} Y^*_{LM}|\bar\nu'>}
   {E_0 - |e_\nu-e_F| - |e_{\nu'}-e_F| - \hbar\omega_{Ln}},
\label{VIND}
\end{equation}
where $L$, $M$ and $n$ denote the quantum numbers of the exchanged collective
vibrations, $f_{Ln}(r) = \beta_{Ln} R_0 (dU/dr)$ 
is the associated radial formfactor
($\beta_{Ln}$ is the deformation parameter associated with the $n-$th mode
of multipolarity $L$, $R_0$ is the ground-state radius and $U$ is the 
average potential), $\hbar \omega_{Ln}$ is the 
vibrational energy of the $n-$th mode, while $e_{\nu}$ and $e_F$ are the
single-particle and Fermi energies. $E_0$ is the pairing correlation energy 
per Cooper pair, which is of the order of $-\Delta$. In practice
we have used $E_0$= - 2 MeV. The pre-factor 2 in Eq.~(\ref{VIND}) comes from 
the two possible time orderings associated with the one-phonon exchange. 

In Fig. \ref{fig_qm} we show the value of the diagonal 
matrix elements  
$\langle \nu \bar \nu | v_{ind}| \nu \bar \nu \rangle$ 
of the induced interaction for the nucleus
$^{120}$Sn as a function of the single-particle energy.
The single-particle levels entering Eq. (\ref{VIND}) 
have been calculated making use of a standard
Saxon-Woods potential, whose parameters are provided in the caption 
of Fig. \ref{fig_qm}. In most of  our calculations we use this potential
and an associated effective mass $m_k$ equal to the bare mass. In the last
section we shall use instead an effective mass $m_k=0.7m$, which
simulates the typical outcome of a Hartree-Fock calculation (with, e.g., 
the Gogny interaction). In those cases, the Woods-Saxon parameters are
changed in such a way that the Fermi energy remains close to the 
experimental value ($V_0^{'}$ $\approx (m/m_k) V_0$ \cite{mahaux},
see e.g. Fig.~\ref{fig_m07}).
The parameters
$\beta_{Ln}$ and $\hbar \omega_{Ln}$ have been determined by
diagonalizing a separable multipole-multipole interaction in the
quasiparticle-random phase approximation (QRPA) adjusting the coupling constants
so as to reproduce the energies and transition probabilities of the
lowest-lying states of each spin and parity. Also shown in Fig. 1 are
the diagonal matrix elements
of the 
Gogny interaction. Here, and in the rest of the paper, we shall employ 
the D1 parametrization.

The matrix elements of $v_{ind}$, which display an average value equal to
-0.15 MeV, are peaked at the 
Fermi energy, within an energy range of essentially 5 MeV
around $e_F$. This behaviour reflects the fact that 
$\langle \nu' \bar \nu' | v_{ind} | \nu,\bar \nu
\rangle$ is controlled, through the energy denominator appearing in
Eq.(\ref{VIND}), by the energy of the low-lying collective vibrations
($\hbar \omega_L(n) < $ 5 MeV). Outside this energy range, the values of 
$\langle \nu' \bar \nu' | v_{ind} | \nu,\bar \nu
\rangle$ become very small, less than 50 keV in absolute value. 
The matrix elements of $v_{Gogny}$ decrease on average smoothly in magnitude 
all the way
from the deepest levels up to the continuum threshold, displaying a value
of about -0.3 MeV at $e_F$.

The induced interaction matrix elements depend
on specific properties of finite nuclei, namely  
the single-particle quantum numbers and energies, and the energy, 
the zero-point amplitude and the transition density of the vibrational
states. 
To study the global
features of the bare, of the induced, and of the Gogny interactions as well as 
to provide a more transparent form for their
matrix elements, we shall present, in the rest of this section, 
also the matrix elements obtained using
the semiclassical Thomas-Fermi (TF) approximation. In fact,
this approximation
averages out shell
effects typical of finite nuclei and provides the overall
energy
dependence of the matrix elements.
Within this context we follow 
Ref.~\cite{vinnas} where semiclassical
expressions for two-body matrix elements have been
presented.
To obtain a smooth behaviour of the diagonal matrix elements as a
function of the continuous energy variable $E$, we compute the Fourier transform
of a two-body interaction $v(\vec r_1,\vec r_2)$ using plane
waves of momenta $\vec p_1$ and $\vec p_2$. The average 
of the quantal matrix elements is 
carried out using the normalized semiclassical density matrix,
\begin{equation}
   \rho_E = {1\over g(E)} \delta(E-H_{cl}) =
   {1\over g(E)} \delta(E-{p2\over {2m_k}}-U(r)).
\label{dos}
\end{equation}
In the above relation 
%$U$ is the nuclear average potential, 
%while 
the semiclassical level density $g$ is
\begin{equation}
   g(E) = {1\over \pi} \int_0^{R_c} dr r^2 ({2m_k\over \hbar^2})^{3/2}
   \sqrt{E-U(r)},
\label{gg}
\end{equation}
$R_c$ being the classical turning point.
We assume a constant value for the effective mass $m_k$, neglecting
its possible dependence on position.
The expression for the semiclassical diagonal matrix
element of $v(\vec r_1, \vec r_2)$ as a function of the 
single-particle energy, is consequently~\cite{vinnas} 
\begin{eqnarray}
&& v(E) = \int d^3r_1 d^3r_2 v(\vec r_1, \vec r_2) \times
% \frac{f_L(r_1) Y_{LM}(\hat r_1)
%   f_L(r_2) Y^*_{LM}(\hat r_2)} {E_0 - 2|E-e_F| - \hbar\omega_L}
 \nonumber\\
&& \times \int {d^3p_1 d^3p_2
   \over (2\pi\hbar)^6}
   {e^{{i\over\hbar} (\vec p_1-\vec p_2)\vec s} \over g^2(E)}
   \delta(E-{p_1^2\over {2m_k}}-U(r_1)) \delta(E-{p_2^2\over {2m_k}}-U(r_2)).
\label{veve}
\end{eqnarray}
For an interaction that depends only on the magnitude of the
relative coordinate $s\equiv \vec r_1-\vec r_2$, like the Gogny 
interaction in the S=0, T=1 channel,
one can integrate first
over $d^3 s$ and then over the momenta $d^3p_1$ and $d^3p_2$, obtaining
the expression
\begin{equation}
v(E)= {c(E)}\int d^3R (E- U(R)) \theta(E-U(R)) v(k_E(R),k_E(R)),
\label{simple}
\end{equation}
where $R$ indicates the center of mass, $v(k,k)$ is the  on-shell
pairing  matrix element of the interaction $v(s)$ 
in the $^1S_0$ channel and in a plane wave basis,
evaluated 
at the local momentum $k = k_E(R)= \frac{1}{\hbar}
\sqrt{2m_k (E- U(R))}$, and $c(E)$ denotes the quantity
 \begin{equation}
  c(E)=\frac{2 {m_k}^3}{4 \pi^4\hbar^6 g^2(E)}.
\label{Jacobian}
\end{equation}
In the next section we shall calculate the pairing gap associated with
the different interactions, and therefore
we shall need the matrix elements between all pairs of particles moving in
time reversal. For this purpose we shall employ a formula analogous  to 
that given in Eq.
(\ref{simple}), namely
$$ v(E,E')=
\sqrt{c(E)c(E')}\int d^3R
\sqrt{(E- U(R))}\sqrt{(E'- U(R))} \times $$
\begin{equation}
\times \theta(E-U(R))\theta(E'-U(R))  v(k_E(R),k_{E'}(R)).
\label{offdiag}
\end{equation}
In this relation $E$ and $E'$ indicate the energies of the single-particle states
of each pair.  

We now turn to the case of the induced interaction, associated with
the matrix elements of Eq. (1).
%, which can be written as
%\begin{equation}
%   v_{ind} (\vec r_1, \vec r_2; E) =
%   2 \sum_{LM} \frac{f_L(r_1) Y_{LM}(\hat r_1) f_L(r_2) Y^*_{LM}(\hat r_2)}
%   {E_0 - 2|E-e_F| - \hbar\omega_L}.
%\label{VIND2}
%\end{equation}
In this case the matrix elements 
depend separately on $\vec r_1$ and $\vec r_2$.
We then integrate over the momenta first, obtaining 
\begin{equation}
   v_{ind}(E) = 2\sum_{nLM} c(E)
   \frac{\int d^3r_1 d^3r_2 f_{Ln}(r_1) Y_{LM}(\hat r_1)
   f_{Ln}(r_2) Y^*_{LM}(\hat r_2) j_0^2(ks) (E-U(R))}
   {E_0 - 2|E-e_F| - \hbar\omega_{nL}},
\label{Vsemi}
\end{equation}
where $j_0$ is the Bessel function and $k\equiv \sqrt{{2m_k\over\hbar^2}(E-U(R))}$.
The index $n$ labels the different phonons of multipolarity $L$ (the phonons
included in the present calculation are specified at the end of this Section).
Note that one can obtain numerical results from Eq. (\ref{Vsemi})
by avoiding the
six-dimensional integration. This is achieved by performing a multipole
expansion of the quantity
\begin{displaymath}
j_0^2(ks) (E-U(R)) = \sum_{lm} {4\pi\over 2l+1} F_l(r_1,r_2;E)
Y_{lm}(\hat r_1) Y_{lm}^*(\hat r_2),
\end{displaymath}
which leads to
\begin{equation}
   v_{ind}(E) = 2\sum_{nL}{4\pi c(E)}
   \frac{\int dr_1 dr_2 r_1^2 r_2^2 f_{Ln}(r_1)
   f_{Ln}(r_2)F_L(r_1,r_2;E)}
   {E_0 - 2|E-e_F| - \hbar\omega_{Ln}}.
\label{Vsemi1}
\end{equation}

In Ref.~\cite{vinnas} the reliability
of the TF approximation in reproducing
the
quantum mechanical (QM) matrix elements
has been checked in the particular case of the pairing matrix elements
of a $\delta$-interaction acting among particles moving 
in a harmonic oscillator (HO)
potential. Note that the TF approach implies an
averaging over the quantum numbers associated with the different
states belonging to the major shells and displaying an energy 
$E_N=(N+{3\over 2})\hbar\omega$.
The TF matrix elements have been compared with the
averaged QM matrix elements (taking degeneracies into account).
Overall agreement was observed (cf. Table II of
Ref.~\cite{vinnas}). 

A similar comparison for
the pairing matrix elements of the Gogny force 
in the $^1S_0$ channel 
is reported in Fig. \ref{fig_semicl}.
% for $m_k=m$. 
It is seen that the TF approach provides a
good approximation to the average quantum matrix elements.
However, the QM diagonal matrix elements, in particular
those associated with $s$-states, are systematically larger
(in absolute value) than the non-diagonal ones. 
One should be aware of the fact that the semiclassical density
matrix can be considered to be the analogue to the one obtained
from a Strutinsky smoothing. Therefore, Eq. (\ref{dos}), 
implicitly, represents a function of about 1$\hbar\omega$ width. 
Thus, the matrix element in Eq. (\ref{veve}) contains 
at the quantum level cross terms corresponding to at least
one major shell. Taking for example the 2s-1d shell, we have
to weigh the two diagonal matrix elements with the 2s and 1d
wavefunctions with a factor 1 and 25, respectively. 
The non-diagonal 2s-1d element obtains the weight factor 10. 
Performing the arithmetic average yields the crosses of 
Fig. \ref{fig_semicl}. In this way, we see that the TF expression
(\ref{veve}) reproduces very well the quantal average. 
The more realistic
case of the Gogny matrix elements in the Woods-Saxon potential
(including spin-orbit) is illustrated in Fig.~\ref{fig_ws}.
The general pattern is similar to that shown in the previous figure.

%spin-orbit now for simplicity), we can see that the TF curve
%goes through the average QM results, but that it sistematically
%underpredict the diagonal ones (in absolute value) by about a factor of 2.
%Only when the off-diagonals are considered an agreement between
%TF and the aveage QM results is possible,
%(for plotting the off-diagonals, the average energy E=(ej+ej')/2
%has been used).

Having assessed its validity, we now employ the semiclassical approximation
to compare the matrix elements of the induced interaction, the effective
Gogny interaction and a bare nucleon-nucleon interaction.
Concerning the latter, the matrix elements of bare nucleon-nucleon 
interactions which contain a repulsive core display a qualitatively different
momentum dependence than that displayed by
the Gogny interaction. We shall instead consider here 
the $v_{low-k}$ interaction, which has been devised to reproduce 
the low-momentum properties
of the nucleon-nucleon force, including the experimental phase shifts,
without the introduction of a repulsive core \cite{Kuo}. In particular, we
shall employ a parametrization of $v_{low-k}$ devised 
to reproduce the properties of
$v_{14}$ Argonne potential. This parametrization requires  a momentum cutoff 
$k_{cut}$ equal to 2.1 fm$^{-1}$.
The matrix elements of $v_{low-k}$ in a plane wave basis have been 
calculated in Ref.~\cite{sedrakian} and are reported in
Fig.~\ref{figure_sedra}.  
It is seen that, while $v_{Gogny}$
is less attractive than $v_{low-k}$ for small momenta, it becomes 
more attractive for $k$ larger than $\approx$1 fm$^{-1}$. 

The semiclassical matrix elements of the Gogny D1 interaction 
(Eq.~(\ref{simple}))
and of the $v_{low-k}$
interactions calculated as a function of the single-particle 
energy for the
nucleus $^{120}$Sn, are displayed in Fig.~\ref{fig_lowk1}. 
We remark that, close to the Fermi energy, 
the matrix elements of the Gogny interaction calculated with
the D1S parametrization are about 10 keV less attractive.
It is seen that the 
general trend is the same as that obtained in infinite matter, with the
difference that the matrix elements of the two interactions look 
somewhat more similar to one 
another. Insight into this difference can be obtained
with the help of  
Eq.(\ref{simple}). This relation is
essentially an average of $v(k)$ calculated at the various local
momenta in the nucleus. From Fig. \ref{figure_sedra} we see that
in the interior  of the nucleus (at larger local momenta) 
$v_{Gogny}$ is more attractive
than $v_{low-k}$, but on the
surface it tends to be less attractive than $v_{low-k}$.

The semiclassical matrix elements of the induced interaction are also 
shown in Fig.~\ref{fig_lowk1}.
These matrix elements are  
strongly peaked around the Fermi
energy and reproduce rather well the average of the 
quantal results already shown  
in Fig.~\ref{fig_qm}. In Fig. \ref{fig_ind1}
we have separated the contributions 
arising from the exchange of phonons of multipolarities $L$=2, 3 and 4.
We include all the phonons with energy below 30 MeV in the calculation. 
The importance of the
contributions with $L$=2 and 3 can be understood in terms of the 
high collectivity and low energy of the associated collective modes.
The energies and deformation parameters of the low-lying modes are
$\hbar \omega_2 = 1.17 $ MeV , $\hbar \omega_3 = 2.42$ MeV, $\hbar \omega_4 =2.47$
MeV and 
$\beta_2 = 0.12 $, $\beta_3 = 0.15$, $\beta_4 = 0.07$.  

We also note that at the Fermi energy the matrix elements of $v_{ind}$ 
are rather large, of the order of those  associated with
the bare nucleon-nucleon potential $v_{low-k}$ (in this connection,
cf. the estimate in ref. \cite{BMII}) .

%are collected in Fig.
%\ref{fig3}, and they are compared to those of the Gogny force in
%\ref{fig4}

%From Fig.~\ref{fig5} one can see that the semiclassical
%treatment, giving the
%correct order of magnitude for the matrix elements, confirms
%the importance of the attractive contribution provided by the
%induced interaction around the Fermi energy. It also gives
%insight into the result of Ref.~\cite{epj}, where it has been
%shown that the induced interaction was essential to bring
%the pairing gap from about one half of the experimental value
%(this is the result obtained with the bare force) to a
%satisfactory agreement.

\section{Pairing gaps}

%In Fig.~\ref{fig_lowk1} we show, together with the matrix elements
%of the Gogny and of the induced interaction, 
%the semiclassical matrix elements associated with
%the low-momentum bare nucleon-nucleon interaction $v_{low-k}$ 
%\cite{Kuo}
%obtained inserting in Eq. (\ref{simple})
%the matrix elements of the Fourier transform of this
%interaction in infinite matter. 

In this section, we shall discuss the results of calculations of the pairing 
gap
which employ the matrix elements discussed above. 
%There is no need to discuss the standard BCS equation for infinite
%matter\ \cite{effmass}. 
In the  
nucleus under study, namely $^{120}$Sn, we obtain the state-dependent
pairing gaps by solving the generalized BCS equations \cite{esb}:   
this means that we solve the HFB equations by treating the 
pairing sector self-consistently, while the mean field is 
described in terms of the Woods-Saxon potential already
used in the calculation of the matrix elements. 
In this way, we take into account scattering processes
between nucleons lying on orbits having
different number of nodes/energies \cite{dd},
associated with non-diagonal matrix
elements of the type $v(E_1,E_2; E'_1, E'_2)$,
with different energies $E_1$ and $E_2$ (or $E'_1$ and $E'_2$). 
This is at variance with the usual 
BCS method in which the pairing gap receives
contributions only from scattering processes between time-reversal 
states.
We shall refer in the following either 
to quantal or to semiclassical calculations; in the latter case, this is
done by solving the same equations as in the quantal case,
but replacing 
the QM matrix elements with the TF ones, using Eq. (\ref{offdiag}) 
and the prescription
\begin{equation} 
E = (E_1 + E_2)/2, \quad  E' = (E'_1 + E'_2)/2.
\label{offdiag1}
\end{equation}
 
%In this case, the HFB equations require 
%involving $J^\pi=0^+$ pairs on
%orbitals which are not in time reversal, but are 
%associated 
 The positive energy states are obtained by setting the
system in a spherical box. We have checked that convergence is 
achieved using $R_{box}$=12 fm and including all the states from
the bottom of the potential up to the positive energy  $E_{cut}$=50 MeV. 

%in the two
%pairs. 

%For further details about the numerical calculations, 
%we refer to the appendix A.  

In infinite neutron matter, the pairing gaps calculated with the interactions
$v_{low-k}$ and $v_{Gogny}$ are very similar up to Fermi momenta
of the order of 0.6 fm$^{-1}$, becoming 
increasingly different from each other at higher
momenta, as can be seen from Fig. \ref{fig_deltaf}
(cf. Ref.~\cite{dds,sedrakian}, where the D1S parametrization of the Gogny 
interaction has been used).
The pairing gap obtained with $v_{low-k}$ 
goes to zero, around saturation, 
as a function of $k_F$ much faster than that associated with 
$v_{Gogny}$; it can also be seen that the pairing gap obtained with
$v_{low-k}$  reproduces quite accurately the result 
found allowing particles to interact through the 
Argonne $v_{14}$ bare N-N potential \cite{schwenk}. 
%In order to assess the dependence on the
%effective mass, results are shown using both $m_k = m$ and $m_k = 0.7 m$.

In the case of $^{120}$Sn,
the pairing gap with the $v_{14}$ Argonne interaction 
and with $m_k=m$,  was calculated previously 
in ref.  \cite{physlett}, obtaining a value of about
2.2 MeV close to the Fermi energy. We have verified that essentially the same
value is obtained with $v_{low-k}$, in keeping 
with the infinite matter case (cf. Fig. \ref{fig_deltaf}). 
In the following, however, we shall adopt the value $m_k= 0.7 m$,
which is the typical value associated with Hartree-Fock calculations
performed with Gogny or Skyrme forces.
In Fig. \ref{fig_m07} we compare the state-dependent pairing gap obtained with
the Gogny force and with the $v_{low-k}$ interaction. We also compare
the pairing gaps obtained inserting the semiclassical pairing
matrix elements in the HFB equations. 
%The accuracy of the semiclassical
%calculation can be assessed in the case of the Gogny force from
%the results displayed in thei nset of 
%and using TF matrix elements, as can be seen in the inset of    
%Fig. \ref{fig_m07}.
We remark that the semiclassical approach takes out the scatter from
shell effects which is a desirable feature when one wishes to
reveal generic trends. 

%the difference between the gaps
%calculated with the Gogny and the Argonne interactions is rather
%pronounced, as it was previously discussed 
%In that reference, a value of the effective mass $m_k =$m was used.
%In fact, the value of the gap
%at the Fermi energy in $^{120}$Sn, calculated with the Gogny
%interaction (and for $m_k=m$), 
%is about 2.7 MeV (cf. the inset of Fig.~\ref{fig_m07})
%while that associated with $v_{14}$ is about  2.2 MeV \cite{physlett}.

%We have performed a similar calculation using the interaction
%$v_{low-k}$ instead of $v_{14}$. 
%The resulting pairing gaps are 
%displayed in the inset of Fig.~\ref{fig_m07} and are similar
%to those obtained previously with $v_{14}$, 

% reports  similar results but now using TF matrix elements
%and the $v_{low-k}$ potential. 
%In fact, the gap associated with $v_{low-k}$
%is very similar to that obtained in ref. \cite{physlett} making use of the $v_{14}$
%interaction. 
%Furthermore, t
%The semiclassical approximation reproduces reasonably
%well the gap obtained in a fully quantum calculation.
%in the case of the
%Gogny interaction.

%It is well known that the value of the pairing gap is very sensitive to the
%single-particle level density, namely to the effective mass. 
%A typical value
%of $m_k$ associated with Hartree-Fock calculations, e.g. with the Gogny
%interaction, is 0.7$m$. 
%Accordingly, 
%in Fig.~\ref{fig_m07} we present also results for a Woods-Saxon potential
%associated with $m_k$ = 0.7$m$.
% which is the
%typical value for the mean field resulting from Hartree-Fock
%calculations. Once again, 
The Gogny interaction leads to a 
pairing gap of 1.4 MeV close to the Fermi energy,
in good agreement with
%the HFB calculations of Ref. \cite{Decharge} and with   
the experimental value deduced from the odd-even mass difference. This value
is about two times larger than that obtained with $v_{low-k}$. 
Note also that the pairing gap obtained with
$v_{low-k}$ reproduces the results obtained with the bare $v_{14}$
interaction~\cite{epj}. We remark that using the D1S parametrization of the 
Gogny force, instead of D1, one obtains a gap at the Fermi energy
of about 1.2 MeV.

In Fig.~\ref{fig_ind} we show the state-dependent pairing gap obtained with
the pairing matrix elements of $v_{total}= v_{low-k}+v_{ind}$, 
resulting from the sum  
of the matrix elements of the bare interaction $v_{low-k}$ 
and of the induced interaction $v_{ind}$.
%using again single-particle wavefunctions and energies of a Woods-Saxon 
%potential associated with $m_k=0.7m$.
The average value of the resulting pairing gap $\Delta_{total}$
at the Fermi energy is about 2 MeV.
This value is about 25\% 
higher than that obtained in Ref.~\cite{epj}, where we solved
the Dyson-Gor'kov equation, taking properly into account 
all polarization effects (induced interaction, self-energy and vertex
corrections)~\cite{Terasaki}.

%Thus the sum $v_{low-k}+ v_{ind}$ is considerably more attractive than
%$v_{Gogny}$ in a small range around $e_F$, but somewhat less
%attractive in a wide range of energies (5 MeV $ < |e -
%e_F| < $ 50 MeV), range in which $v_{Gogny}$ leads to
%non-negligible contributions to the pairing gap.

This is consistent with the fact that according to 
Nuclear Field Theory (NFT) \cite{NFT},
if one considers the effects of the exchange of phonons between pairs of nucleons,
one has to consider at the same time processes where the phonon is
absorbed by the same nucleon which has virtually excited it. Such processes
lead to self-energy ($\omega$-mass and single-particle splitting) as well as
to vertex correction phenomena. Of these processes and for the nucleus under
discussion ($^{120}$Sn), $\omega-$mass effects \cite{mahaux} 
are the most important. Taking such
effects into account is equivalent to use a residual interaction
$v_{ren}= v_{total}Z_{\omega}$, where $Z_{\omega} = (m_{\omega}/m)^{-1}$ is the 
quasiparticle strength \cite{mahaux} at the Fermi energy (cf. Appendix A). 
Consequently $\Delta_{ren} = Z_{\omega} \Delta_{tot}$. Because 
$v_{ind}$ is proportional to $\beta_L^2/DEN_L$, that is to the square of the
deformation parameters associated with the low-lying collective vibration
and inversely proportional to the energy denominator $DEN_L$ appearing 
in Eq.(\ref{VIND}), a
typical error of 20\% on $\beta_L$ and $DEN_L$ implies a 60\% error in $v_{ind}$.
Making use of this fact and of the results of Figs. \ref{fig_qm} and
\ref{fig_ind} one obtains,
as discussed in Appendix B, $\Delta_{ren} \approx Z_{\omega} \Delta_{tot}
\approx 1.35 \pm 0.15 $ MeV. It is seen that the 60\% error 
in $v_{ind}$ has been reduced to a 10\% in $\Delta_{ind}$. This is because
the stronger the particle-vibration coupling is, the larger $\Delta_{tot}$ 
but the smaller the amount of single-particle content of levels around
$\epsilon_F$ (and thus the smaller $Z_{\omega}$), and viceversa. The
self consistency between collectivity of the modes, strength of
the particle-vibration coupling, and quasi-particle residue at
the pole, typical of NFT, allows theory to make predictions which are 
more accurate
than the basic parameters entering the calculation, as a result of a 
delicate process of cancellation of errors.

%Because $Z_{\omega} \approx 0.6-0.7$,
%one obtains $\Delta_{ren} \approx Z_{\omega}\times 2 $ MeV 
%$\approx 1.2-1.4 $ MeV in
%overall agreement with the experimental findings.

\section{Conclusions}

Summing up, we have found that the small difference existing between the matrix
elements of the bare $v_{low-k}$ interaction and those of the Gogny
interaction leads to important differences between the pairing gaps
associated with the two forces ($\Delta_{Gogny} \approx 2 \Delta_{low-k}$, for
$m_k \approx 0.7 m$).
This difference is removed and eventually overwhelmed by including the 
attractive contribution coming from the induced interaction $v_{ind}$ arising
from the exchange of  surface vibrations,
which acts only on a rather small energy range around $e_F$. 
We note  that the pairing gap  in $^{120}$Sn is not much changed, 
including also the effect of spin fluctuations \cite{spin}, 
which instead give the dominant  (repulsive) contribution to the 
induced interaction in neutron matter.

%In fact,
%the function $v_{ren} = v_{low-k}+ v_{ind}$ is much more attractive than
%$v_{Gogny}$ in the same energy range, so as to compensate for the
%difference between $v_{low-k}$ and $v_{Gogny}$ in the remaining 40 MeV
%range over which the integration is to be carried out to achieve
%convergence. 
The 25\% excess displayed by the resulting total pairing gap 
with respect to the experimental value is corrected by considering the 
corresponding quasiparticle strength at the Fermi energy.
An important question which remains to be addressed quantitatively is 
the ability $v_{low-k} + v_{ind}$ as well as $v_{Gogny}$ have 
to describe  in detail the isotopic and nuclear structure dependences 
displayed by $\Delta_{exp}$.
This subject will be addressed in a future publication.

Within this context,  we note that the 
dependence of the pairing gaps 
generated by
$v_{Gogny}$ and by $v_{low-k}+ v_{ind}$ on 
temperature 
and rotational frequency  is expected to be quite
different, as $v_{Gogny}$ and $v_{low-k}$ are essentially independent
of these parameters, while $v_{ind}$ is quite sensitive to them.

It is also worth mentioning that a recent  study in symmetric
nuclear matter revealed a very strong extra attraction coming from the induced
interaction \cite{Shen}. Nuclear matter and finite nuclei may, however,
show only qualitative similarities in this case and a quantitative link 
may be difficult to establish.

%While the above results provide an overall understanding of the mechanism which
%is at the basis of the detailed microscopic results leading to
%$ \Delta_{bare+ind} \approx \Delta_{Gogny}$, they are just
%qualitative, as they are based only on the diagonal
%matrix elements.

%Finally, let us note that while $v_{Gogny}$ is able to produce values
%of the pairing gap which are in overall agreement with the experimental
%findings, it has been found~\cite{PRL} that $v_{low-k}+ v_{ind}$ is able to
%reproduce the local $A-$dependence of the pairing gap associated with the
%dynamics of the shell structure. 

\vspace{1cm}

We thank L. Coraggio for providing us with the matrix elements of the 
$v_{low-k}$ interaction.

\newpage 

\section*{Appendix A}

\renewcommand{\theequation}{A.\arabic{equation}}
\setcounter{equation}{0}
\underline{Simple estimate of the pairing gap}

\vskip 1cm

In what follows we consider some of the consequences the particle-vibration coupling has on the pairing correlations of particles moving in a single j-shell interacting through a 
bare nucleon-nucleon pairing potential with constant matrix elements $G$ \cite{Brink}.

For this simple model, 
the value of the occupation numbers $U_{\nu}$ and $V_{\nu}$ must be the same for all the 
$2j+1$ orbitals. In particular, the occupation probability for the case when the system
is occupied with $N= \Omega$ particles (half filled shell), where
\begin{equation}
\Omega = \frac{2j+1}{2}
\end{equation}
is
\begin{equation}
 V= \sqrt{\frac{N}{2 \Omega}} = \sqrt {\frac{1}{2}},
\end{equation} 
\begin{equation}
 U= \sqrt{1 - \frac{N}{2 \Omega}} = \sqrt {\frac{1}{2}}.
\end{equation} 
Consequently, the pairing gap is given by the relation
\begin{equation}
\Delta= G \sum_{\nu>0} U_{\nu}V_{\nu} = \frac {G \Omega}{2}. 
\end{equation}
Because the density of levels is proportional to the $\omega-$mass ($m_{\omega}$),
the effective (dressed) degeneracy can be written
\begin{equation}
\Omega_{eff} = \frac {\Omega}{Z_{\omega}},
\end{equation}
in terms of the quasiparticle strength at the Fermi energy  
${Z_{\omega}}= (m_\omega/m)^{-1}$ where $m_{\omega}= m(1+ \lambda)$,
$\lambda$ being the mass enhancement factor. In the  case of nuclei
this dimensionless quantity (which measures the strength with which
nucleons couple to low-lying collective vibrations)  
is of the order of 0.5, closer to the strong than to the weak
coupling situation of BCS (in which case $\Delta \sim \lambda$) \cite{Schrieff}. 

Because of their coupling to vibrations, nucleons spend part of the time in more 
complicated 
configurations than pure single-particle states. 
The quantity (quasiparticle strength \cite{mahaux})
$Z_{\omega}= (1+ \lambda)^{-1} \approx 0.7$ measures the content
 of single-particle strength present in levels around the Fermi energy 
 available for nucleons to interact through a two-body interaction,
 in particular through through a pairing
force. Consequently, due to the self-energy processes arising from 
the particle-vibration
coupling phenomenon, the pairing strength becomes $G {Z_{\omega}}^2$. 

The exchange of vibrations between pairs of nucleons moving in  time reversal states
close to the Fermi energy gives rise to an effective pairing 
interaction of strength $g_{p-v}{Z_{\omega}}^2$, where $g_{p-v}$ stands for the 
induced pairing interaction controlled by the particle-vibration coupling vertices. Taking these effects into
account,
one can write
\begin{equation}
\Delta= \frac{Z_{\omega}}{2} (G + g_{p-v}) \Omega 
%\approx 0.7 \frac{1}{2}(G+ g_{p-v})\Omega.
\label{the16}
\end{equation}
Identifying $G$ with $v_{low-k}$ and $g_{p-v}$ with $v_{ind}$, 
one finds from Fig.~\ref{fig_ind} that
\begin{equation}
\frac{1}{2} (G+g_{p-v})\Omega \approx  {\rm 2 MeV}.
\end{equation}
As seen from Eq.~(\ref{the16}), the number to be compared with experimental findings is
\begin{equation}
\Delta_{ren} = Z_{\omega} \times 2 {\rm MeV} \approx 1.4 {\rm MeV}.
\end{equation}
Note that in the above discussion we have not considered the errors to be
expected in estimating both $g_{p-v}$ and $Z_{\omega}$. This subject is 
taken up in Appendix B. 

\newpage

\section*{Appendix B}

\renewcommand{\theequation}{B.\arabic{equation}}
\setcounter{equation}{0}
\underline{Simple estimate of $\lambda$ and $Z_{\omega}$}

\vskip 1cm

\begin{figure}
\begin{center}
\includegraphics[scale=0.6]{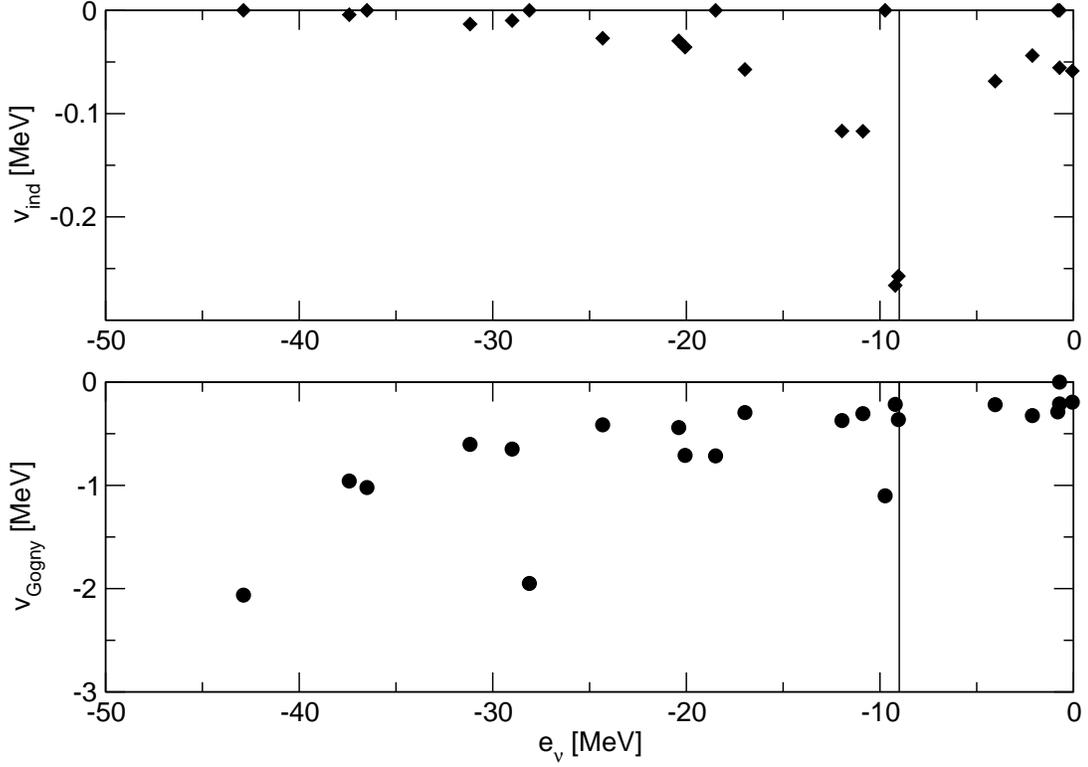}
\end{center}
\caption{The nucleus $^{120}$Sn. Diagonal pairing matrix elements of the induced interaction
(upper panel, solid diamonds) and of the Gogny force 
(lower panel, solid circles),
displayed as a function of the single-particle energy $e_{\nu}$ of
the state $\nu$ calculated using the bare nucleon mass and the single-particle wavefunctions  of 
a Woods-Saxon potential with standard parameters (depth $V_0 $= -49 MeV, diffusivity
$a=$0.65 fm, and radius $R_0=$ 6.16 fm), including the spin-orbit term,
parametrized according to ref. \cite{BMI}. 
Also shown by means of vertical lines is the
position of the Fermi energy, $e_F = -9.1 $ MeV. Note the different scale in the two figures.}
\label{fig_qm}
\end{figure}

\begin{figure}
\begin{center}
\includegraphics[scale=0.6,angle=-90]{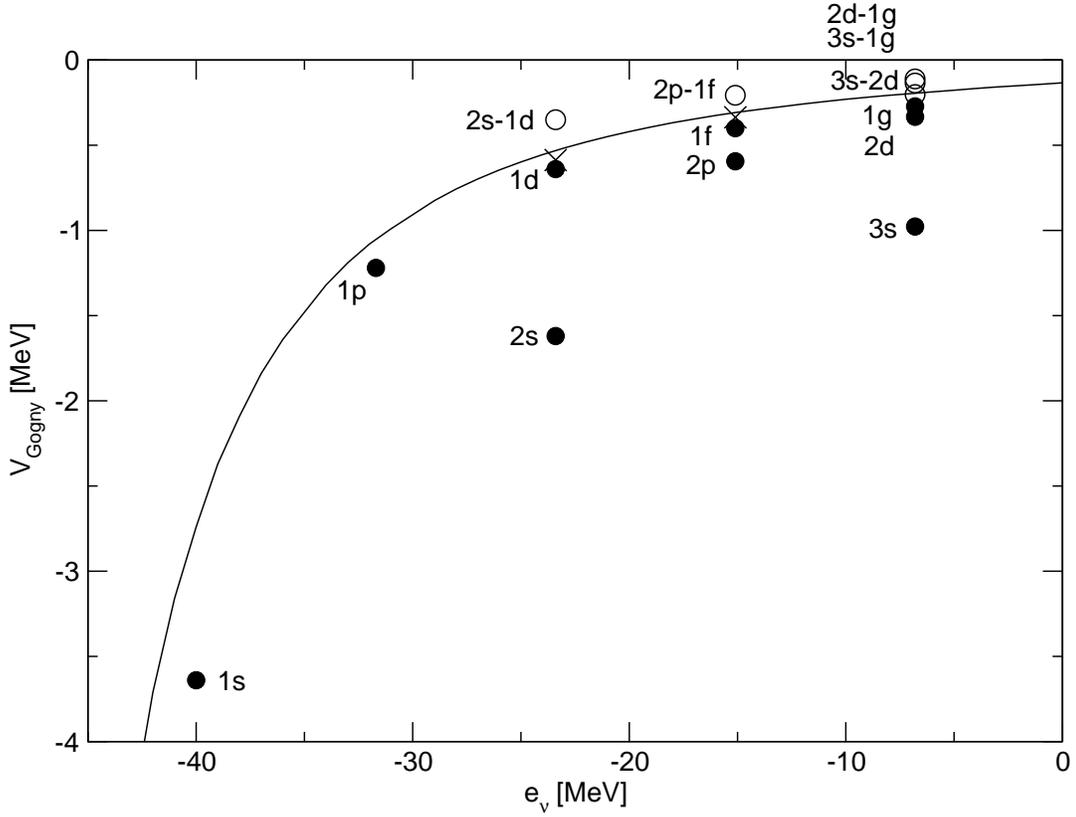}
\end{center}
\caption{Schematic model for the nucleus $^{120}$Sn.
The semiclassical pairing matrix elements of
the Gogny force (solid line) are compared to the quantal
matrix elements for the
case in which the single-particle wavefunctions have been calculated making
use of a harmonic oscillator potential without the spin-orbit term
(and $m_k = m$). The diagonal and
non-diagonal matrix elements are denoted by filled and open
circles respectively. The crosses correspond to the weighted averages of
the matrix elements within a shell.}
\label{fig_semicl}
\end{figure}

\begin{figure}
\begin{center}
\includegraphics[scale=0.6]{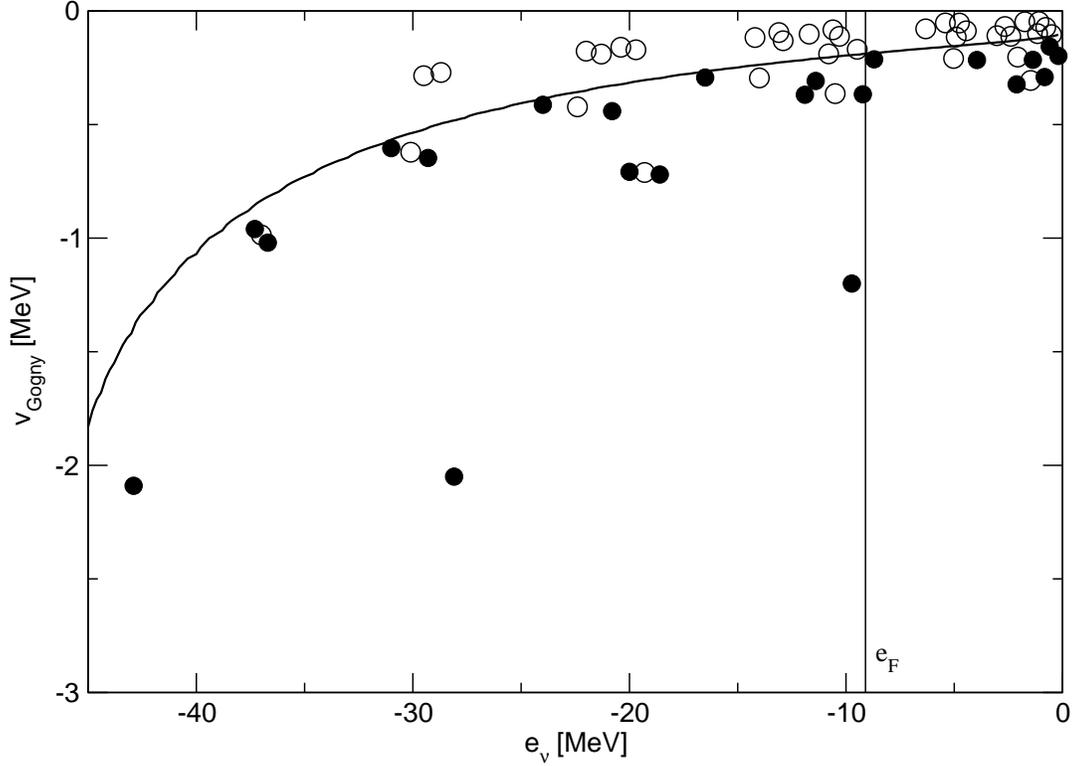}
\end{center}
\caption{
The nucleus $^{120}$Sn. The same as Fig.~\ref{fig_semicl} 
(i.e. also with $m_k = m$) but for the fact that the
single-particle wavefunctions have been calculated making use of the 
Woods-Saxon potential including the spin-orbit term already used
for Fig.~\ref{fig_qm}, where the diagonal matrix elements 
have already been shown. The non-diagonal
matrix elements are plotted here at an energy $e_\nu$ which is the average
between the energies of the initial and final states.}
\label{fig_ws}
\end{figure}

\begin{figure}
\begin{center}
\includegraphics[scale=0.6,angle=-90]{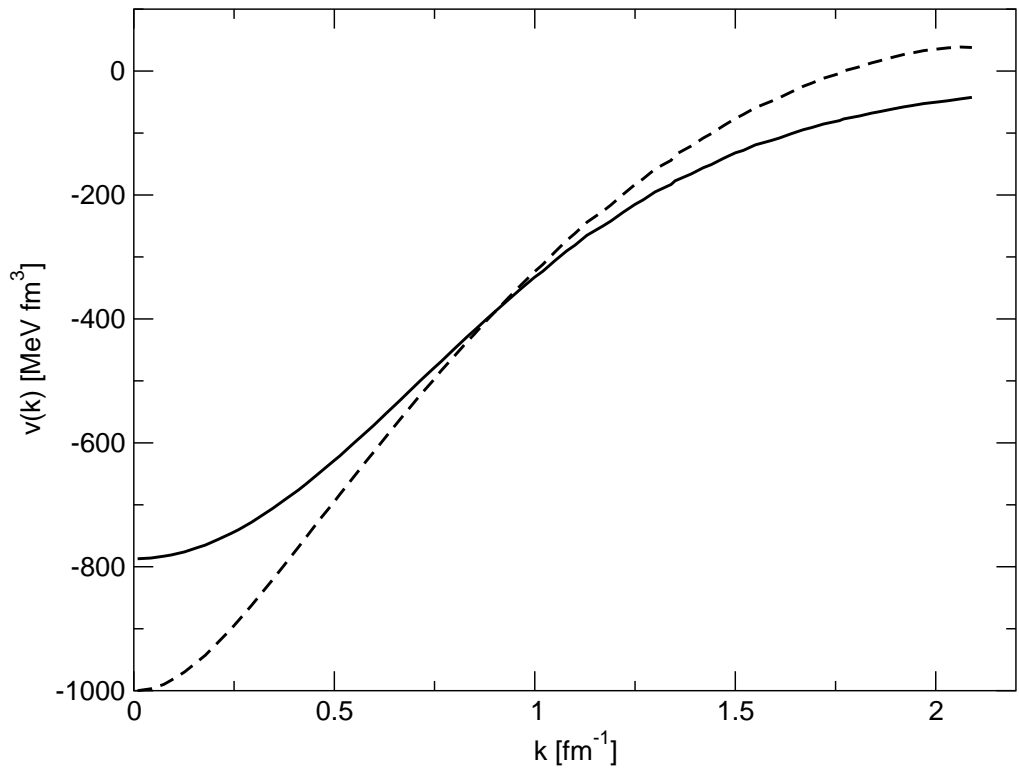}
\end{center}
\caption{The matrix elements of the Gogny (solid curve) and of the $v_{low-k}$ 
(dashed curve) interactions 
are plotted as a function of momentum.}
\label{figure_sedra}
\end{figure}

\begin{figure}
\begin{center}
\includegraphics[scale=0.6]{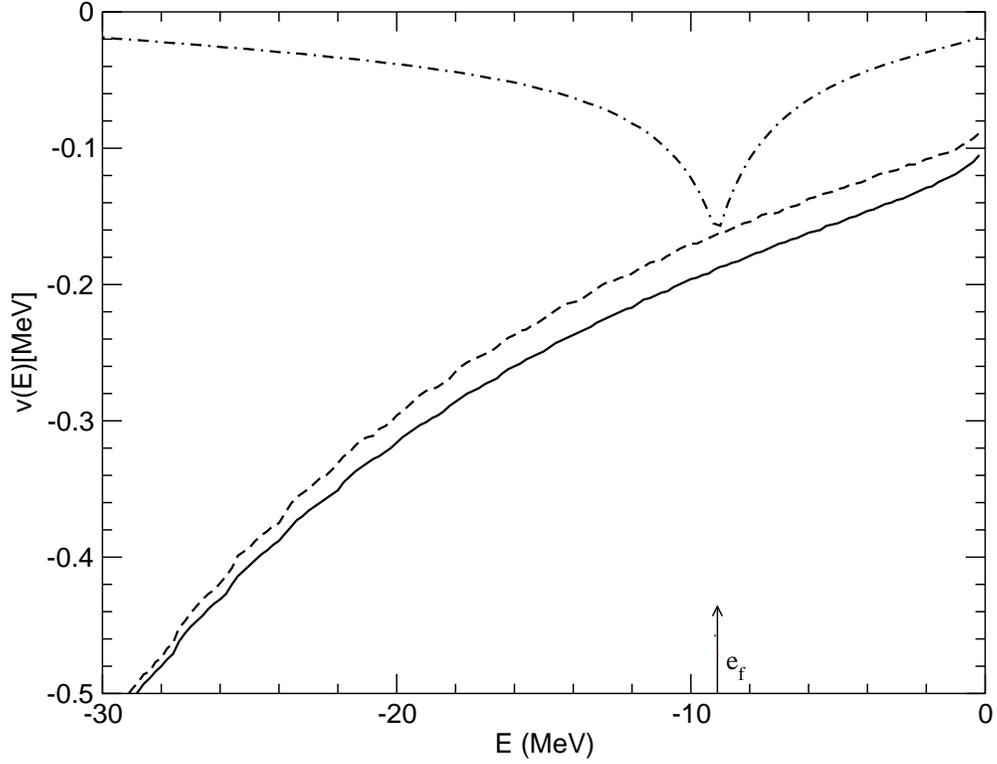}
\end{center}
\caption{
The nucleus $^{120}$Sn. The semiclassical matrix elements 
of the induced
interaction, calculated according to Eq. (\ref{Vsemi1})(dash-dotted curve), are
compared with the matrix elements of the Gogny force (solid curve,
cf. Fig.~\ref{fig_ws}) and with those of the $v_{low-k}$
interaction (dashed curve). Calculations are performed with $m_k=m$ 
and with the same Woods-Saxon potential used in Figs. \ref{fig_qm} and 
\ref{fig_ws}.}
\label{fig_lowk1}
\end{figure}

\begin{figure}
\begin{center}
\includegraphics[scale=0.6]{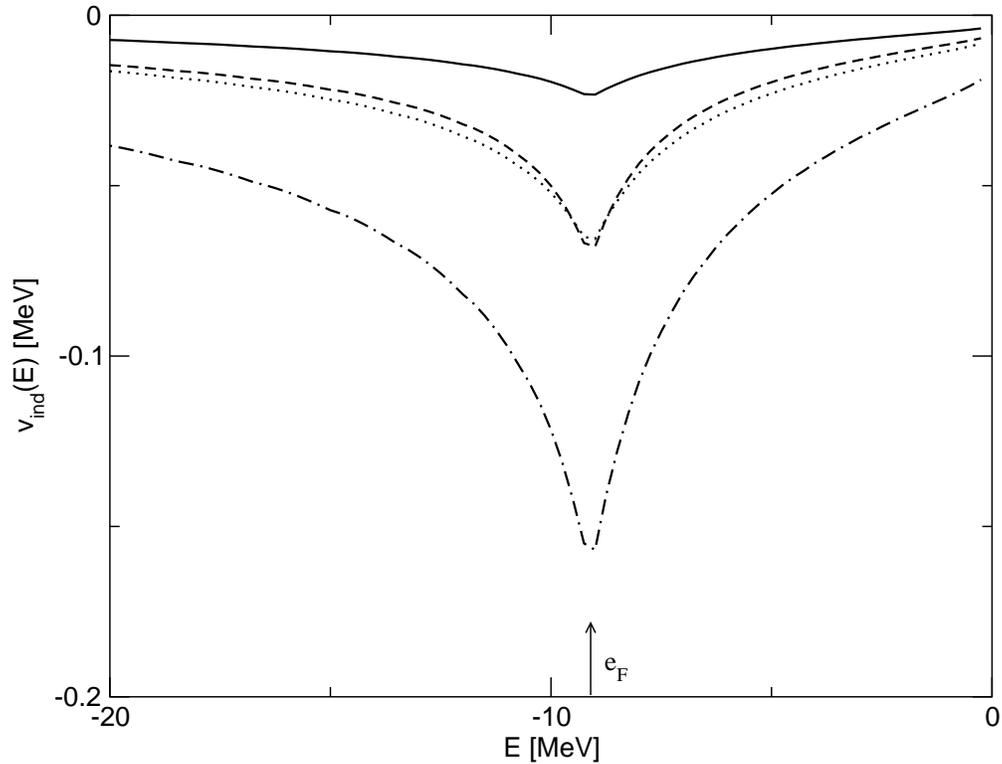}
\end{center}
\caption{The nucleus $^{120}$Sn. The semiclassical matrix elements of
the induced interaction, as a function of the energy of the single-particle
levels
(calculated with $m_k=m$) associated with  the multipolarities
$L=$ 2, 3, and 4 of the formfactor (see Eq.(\ref{VIND})), are shown by a dashed, dotted and
solid line respectively. The sum of the three contributions is displayed
by means of a dash-dotted line (cf. Fig. \ref{fig_lowk1}).}
\label{fig_ind1}
\end{figure}

\begin{figure}
\begin{center}
\includegraphics[scale=0.6]{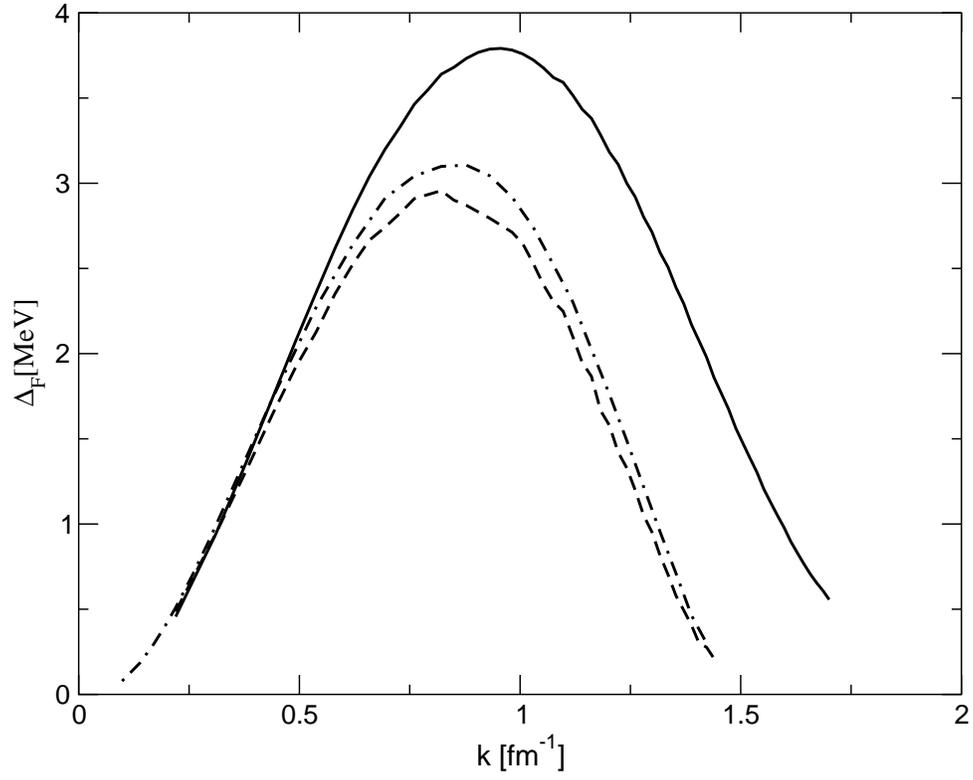}
\end{center}
\caption{Pairing gaps calculated in neutron matter as a function of the Fermi
momentum, obtained with the Gogny interaction (solid line), the Argonne
$v_{14}$ potential (dash-dotted line) and the $v_{low-k}$ potential (dashed
line). The bare effective mass has been used in the calculation.}
\label{fig_deltaf}
\end{figure}

\begin{figure}
\begin{center}
\includegraphics[scale=0.6]{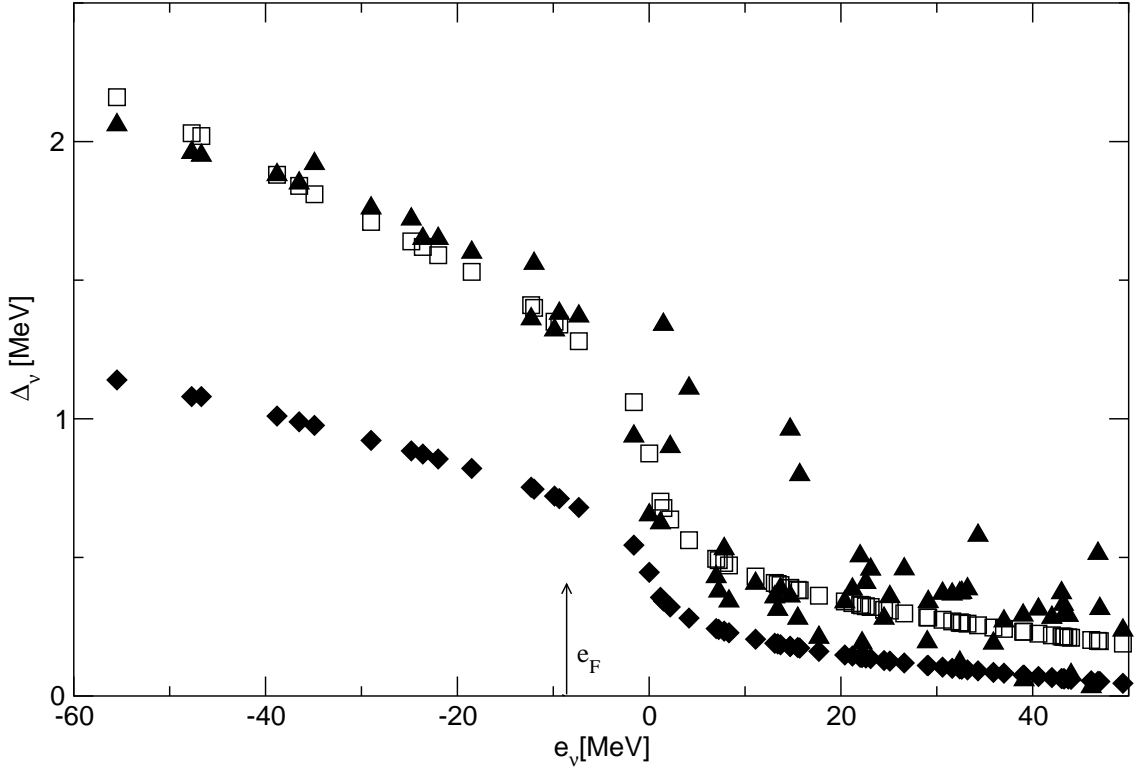}
\end{center}
\caption{
State-dependent pairing gaps of $^{120}$Sn calculated with a Woods-Saxon 
potential
(with depth $V_0$ = -64 MeV, diffusivity $a = $ 0.65 fm and radius $R_0 = $
6.17 fm), as a function of the single-particle energy. 
The $k-$mass $m_k$ was set equal to 0.7$m$.
The Fermi energy
is $e_F = $- 8.6 MeV. Solid triangles (open squares) display the
results of a HFB calculation with the Gogny interaction, 
with quantal (semiclassical) matrix elements.  
The solid diamonds refer instead to a HFB
calculation using the semiclassical matrix elements of the $v_{low-k}$
potential.}
%The inset refers to analogous calculations performed with a Woods-Saxon 
%potential with depth $V_0= -49 $ MeV, associated with an effective mass $m_k= m$ and
%a Fermi energy $e_F = -9.1 $MeV (already used in Fig.~\ref{fig_qm}).}
\label{fig_m07}
\end{figure}

\begin{figure}
\begin{center}
\includegraphics[scale=0.6]{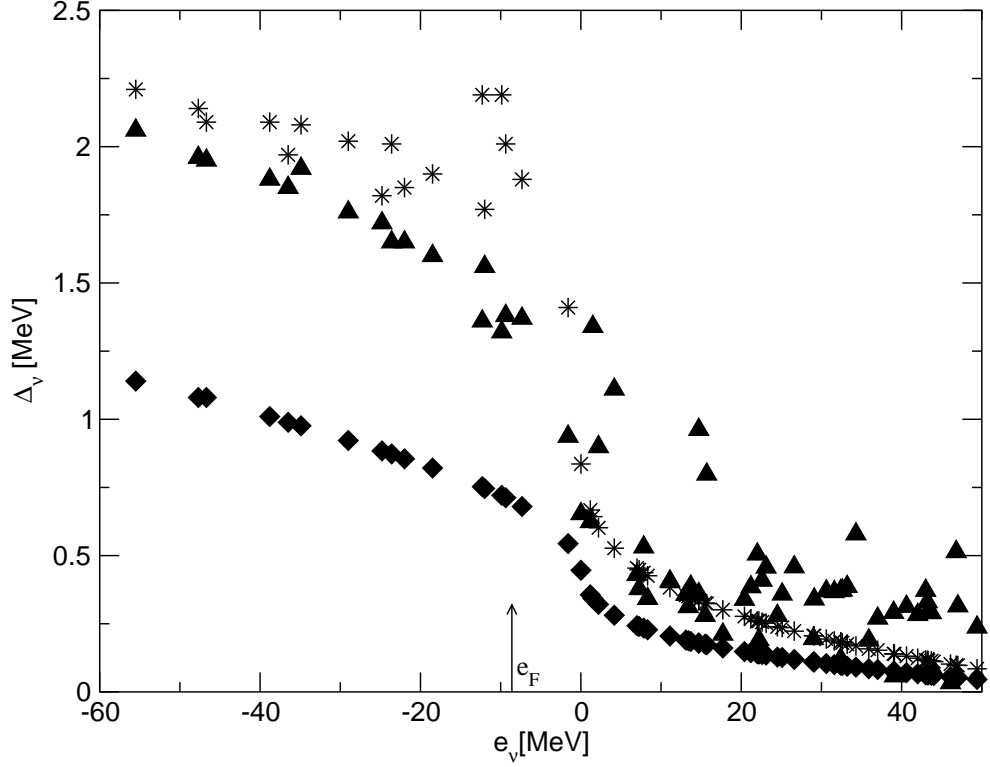}
\end{center}
\caption{State-dependent pairing gap of $^{120}$Sn 
obtained making use of the Woods-Saxon
potential used in calculating the results displayed in Fig. \ref{fig_m07} 
(i.e. with $m_k = 0.7 m$)
and different interactions. Diamonds and 
triangles show again the gaps obtained with the $v_{low-k}$ and $v_{Gogny}$
interactions, respectively. Stars show the gap associated 
with the matrix elements obtained summing the matrix elements of 
$v_{low-k}$ and $v_{ind}$. The latter have been evaluated using Eqs. 
(\ref{Vsemi1}) and (\ref{offdiag1})}.
\label{fig_ind}
\end{figure}

From Fig. \ref{fig_qm} it is seen that the average values of $v_{ind}$ and
$v_{low-k}$ at the Fermi energy are $<v_{ind}>$ = -0.15 MeV and
$<v_{low-k}>$ = -0.16 MeV. Because $v_{ind}$ is of the order  of
$\sum_L \beta_L^2/DEN_L$, the matrix elements of $v_{ind}$ will display 
an error 
equal to twice the average error displayed by $\beta_L$ plus that displayed
by $DEN_L$. Assuming
this error to be 20\% for both quantities (a rather extreme negative situation)
one concludes
that $<v_{ind}> = 0.15 \pm $0.09 MeV. Consequently
$<v_{tot}> = < v_{ind}> + <v_{low-k}> = -0.31 \pm $0.09 MeV.
Because $\Delta_{total}$ = 2 MeV for $< v_{tot}>$ =  -0.31 MeV,
one obtains $\Delta_{total} = (2  \pm 0.6 ) $ MeV.

We now proceed to the calculation of $\lambda = N(0) <v_{ind}>$,
where $N(0)$ is the density of single-particle levels around the 
Fermi energy for one spin orientation and one type of nucleon. 
Making use of the value $N(0) = $ 3.4 MeV$^{-1}$ appropriate for $^{120}$Sn, one
gets $\lambda = 0.5 \pm 0.3$. Consequently $\Delta_{ren} = 
Z_{\omega} \Delta_{tot}$ =
1.35 $\pm $ 0.15 MeV, where 
$Z_{\omega}= (1 +\lambda)^{-1}$. We thus note a conspicuously smaller error
for $\Delta_{ren}$ than for $v_{ind}$. This is because the larger is $\lambda$ 
(induced pairing contribution), the smaller is the content of
single-particle strength  and thus of $Z_{\omega}$, associated with levels
lying close to the Fermi energy, and viceversa.

\newpage

%\begin{figure}
%\begin{center}
%\includegraphics[scale=0.6]{fig5_schuck.eps}
%\end{center}
%\caption{The semiclassical matrix elements of the induced
%interaction (solid line), already shown in Fig~\ref{fig_ind1},
%are compared with the matrix elements of the $v_{low-k}$
%interaction (dashed line).}
%\label{fig_ind2}
%\end{figure}

%\begin{figure}
%\begin{center}
%\includegraphics[scale=0.6,angle=-90]{10.eps}
%\end{center}
%\caption{
%\label{fig_m1}
%\end{figure}

\end{document}